\documentclass[11pt,twoside]{article}
\usepackage{my}
\usepackage{latexsym}
\usepackage{amssymb}
\usepackage{amstext}
\usepackage{bbm}
\usepackage{verbatim}

\usepackage[final]{voronkov}

\makeindex

\newcommand{\paper}{paper}
\newcommand{\xone}[2]{{#1}_1,\ldots,{#1}_{#2}}
\newcommand{\setof}[1]{{\{#1\}}}
\newcommand{\eqref}[1]{(\ref{#1})}

\newcommand{\Index}[1]{\index{#1}}
\newcommand{\DI}[2]{\emph{#1}\Index{#2}}
\newcommand{\DII}[3]{\emph{#1}\Index{#2}\Index{#3}}

\newcommand{\orl}{\vee}
\newcommand{\bigorl}{\bigvee}
\newcommand{\bigandl}{\bigwedge}
\newcommand{\andl}{\wedge}
\newcommand{\notl}{\neg}
\newcommand{\iffl}{\leftrightarrow}

\newcommand{\false}{\bot}

\newcommand{\TA}{\mathrm{TA}}

\newcommand{\nat}{\mathbb{N}}

\newcommand{\KBo}{\succ}
\newcommand{\oKB}{\prec}
\newcommand{\KBe}{\KBo_{\mathit{lex}}}
\newcommand{\KBw}{\KBo_{w}}

\newcommand{\Arith}{C_{\mathit{arith}}}
\newcommand{\Simp}{C_{\mathit{simp}}}
\newcommand{\Triang}{C_{\mathit{triang}}}
\newcommand{\Cha}{C_{\mathit{chain}}}

\newcommand{\arity}{\mathit{arity}}

\newcommand{\atleast}{\mathit{at\_least}}

\newcommand{\Exists}{\mathit{exists}}
\newcommand{\TT}{\mathit{tnt}}
\newcommand{\Con}{\mathit{Con}}

           \title{Knuth-Bendix constraint solving is NP-complete}

\author{
  Konstantin Korovin and %
  Andrei Voronkov \\
  The University of Manchester\\
   \{korovin,voronkov\}@cs.man.ac.uk}

\runningtitle{KBO constraint solving is NP-compete}
\runningauthor{K.\ Korovin and A.\ Voronkov}

\begin{document}

\maketitle

\begin{abstract}
  We show the NP-completeness of the existential theory of term algebras with
  the Knuth-Bendix order by giving a nondeterministic polynomial-time
  algorithm for solving Knuth-Bendix ordering constraints.
\end{abstract}




\mymark

\vspace{2ex}



                      \section{Introduction}

Solving ordering constraints in term algebras with various reduction orders
is used in rewriting to prove termination of recursive definitions and in
automated deduction to prune the search space
\cite{Comon:JFCS:OrderingConstraints:1990,%
Kirchner:CPBT:ConstraintsInDeduction:1995,Nieuwenhuis:CADE:Invited:1999}.
\Cite{Nieuwenhuis:CADE:Invited:1999} connects further progress in
automated deduction with constraint-based deduction.

Two kinds of orders are used in automated deduction: the Knuth-Bendix order
\cite{KnuthBendix:Pergamon:WordProblems:1970} and various versions of
recursive path orders \cite{der82,KaminLevy:LPO:1980}. The Knuth-Bendix order is used in the
state-of-the-art theorem provers, for example, E \cite{Schulz:CADE:E:1999},
SPASS \cite{Weidenbach+:CADE:SPASS1.0.0:1999}, Vampire
\cite{RiazanovVoronkov:CADE:Vampire:1999}, and Waldmeister
\cite{Hillenbrand+:JAR:Waldmeister:1997}. There is extensive literature on
solving recursive path ordering constraints (e.g.,
\cite{Comon:JFCS:OrderingConstraints:1990,%
JouannaudOkada:ICALP:SubtermConstraints:1991,nie93,%
Narendran+:CSL:RPOinNP:1999}). The decidability of Knuth-Bendix ordering
constraints was proved only recently in
\cite{VoronkovKorovin:LICS:KnuthBendix:2000}. The algorithm described in that
paper shows that the problem belongs to 2-NEXPTIME. It was also shown that
the problem is NP-hard by reduction of the solvability of systems of linear
Diophantine equations to the solvability  of the Knuth-Bendix ordering
constraints. In this \paper\ we present a nondeterministic polynomial-time
algorithm for solving Knuth-Bendix ordering constraints, and hence show that
the problem is contained in NP for every term algebra with a Knuth-Bendix
order. As a consequence, we obtain that the existential first-order theory of
any term algebra with a Knuth-Bendix order is NP-complete too. Let us note
that the problem of solvability of a Knuth-Bendix ordering constraints
consisting of a single inequality can be solved in polynomial time
\cite{KorovinVoronkov:RTA:KBorientabilityIsNP:2001}.

This \paper\ is structured as follows. In Section~\ref{sec:preliminaries} we
define the main notions of this paper. In Section \ref{sec:isolated:forms} we
introduce the notion of isolated form of constraints and show that every
constraint can be effectively transformed into an equivalent disjunction of
constraints in isolated form. This transformation is represented as a
nondeterministic polynomial-time algorithm computing members of this
disjunction. After this, it remains to show that solvability of constraints
in isolated form can be decided by a nondeterministic polynomial-time
algorithm. In Section~\ref{sec:isolated->diophantine} we present such an
algorithm using transformation to systems of linear Diophantine inequalities
over the weights of variables. Finally, in Section~\ref{sec:main:result} we
complete the proof of the main result and present some examples. 
Section~\ref{sec:related} discusses
related work and open problems.

                           \section{Preliminaries}
                          \label{sec:preliminaries}

A \DI{signature}{signature} is a finite set of function symbols with
associated arities. In this \paper\ we assume an arbitrary but fixed
signature $\Sigma$. \DI{Constants}{constant} are function symbols of the
arity $0$. We assume that $\Sigma$ contains at least one constant. We denote
variables by $x,y,z$ and terms by $r,s,t$. The set of all ground terms of the
signature $\Sigma$ can be considered as the \DI{term algebra}{term algebra}
of this signature, \DI{$\TA(\Sigma)$}{TA@$\protect\TA(\Sigma)$}, by defining
the interpretation $g^{\TA(\Sigma)}$ of any function symbol $g$ by
$g^{\TA(\Sigma)}(\xone{t}{n}) = g(\xone{t}{n})$. For details see e.g.\
\cite{Hodges:ModelTheory:1993} or \cite{Maher:LICS:AxiomTermAlgebra:1988}. It
is easy to see that in term algebras any ground term is interpreted by
itself.

Denote the set of natural numbers by \DI{$\nat$}{N@$\protect\nat$ --- the set
of natural numbers}. 
The Knuth-Bendix order is a family of orders parametrized by two parameters:
a weight function and a precedence relation.

\begin{definition}[weight function]
  We call a \DI{weight function}{weight function} on
  $\Sigma$ any function $w: \Sigma \rightarrow \nat$ such that (i) $w(a) > 0$
  for every constant $a \in \Sigma$, (ii) there exist at most one unary
  function symbol $f \in \Sigma$ such that $w(f) = 0$. Given a weight
  function $w$, we call $w(g)$ the \DI{weight}{weight!of function symbol} of
  $g$. The \DI{weight}{weight} of any ground term $t$, denoted
  \DI{$|t|$}{$\mid t \mid$ --- weight of $t$}, is defined as follows: for
  every constant $c$ we have $|c|=w(c)$ and for every function symbol $g$ of
  a positive arity $|g(t_1,\ldots,t_n)| = w(g)+ |t_1| + \ldots + |t_n|$.
\end{definition}

 These conditions on the weight function ensure
that the Knuth-Bendix order is a simplification order total on ground terms
(see, e.g., \cite{BaaderNipkow:98}). In this paper, \emph{\DI{$f$}{f@$f$}
will always denote a unary function symbol of the weight $0$}.

The following lemma is straightforward.
\begin{Lemma}
\label{lem:weight:props}%
  Every weight function satisfies the following properties.

  \begin{enumerate}
    \item The weight of every term is positive.

    \item If $\Sigma$ contains no unary function symbol of the weight $0$,
    then for every natural number $n$ there is only a finite number of terms
    of the weight $n$. If $\Sigma$ contains the unary function symbol of the
    weight $0$, then every weight contains either no terms at all or an
    infinite number of different terms.

    \item If a term $s$ is a subterm of $t$ and $|s|=|t|$, then $t$ has the
    form $f^m(s)$ for some $m$ (recall that $f$ is the function symbol of the
    weight $0$). $\QED$
  \end{enumerate}
\end{Lemma}

\begin{definition}
  A \DI{precedence relation}{precedence relation} on $\Sigma$ is any total
  order $\gg$ on $\Sigma$. A precedence relation $\gg$ is said to be
  \DI{compatible}{compatible precedence relation} with a weight function $w$
  if the existence of a unary function symbol $f$ of the weight zero implies
  that $f$ is the greatest element w.r.t.\ $\gg$.
\end{definition}

In the sequel we assume a fixed weight function \DI{$w$}{w@$w$ --- weight
function} on $\Sigma$ and a fixed precedence relation $\gg$ on
$\Sigma$, compatible with $w$.

\begin{definition}
The \DI{Knuth-Bendix order}{Knuth-Bendix order} on $\TA(\Sigma)$ is the
binary relation \DI{$\KBo$}{$\protect\KBo$ --- Knuth Bendix order} defined
as follows. For any ground terms $t = g(t_1,\ldots,t_n)$ and $s =
h(s_1,\ldots,s_k)$ we have $t \KBo s$ if one of the following conditions
holds:

\begin{enumerate}
  \item $|t|>|s|$;\label{cond:KB:1}

  \item $|t|=|s|$ and $g \gg h$;\label{cond:KB:2}

  \item $|t|=|s|$, $g=h$ and for some $1 \leq i \leq n$ we have
  $t_1=s_1,\ldots,t_{i-1}=s_{i-1}$ and $t_i \KBo s_i$.\label{cond:KB:3}
\end{enumerate}
\end{definition}

Note that the Knuth-Bendix order is a total monotonic well-founded order,
see, e.g., \cite{BaaderNipkow:98}.
Some authors \cite{mar87,BaaderNipkow:98} define Knuth-Bendix orders with
real-valued weight functions. We do not consider such orders here, because
for real-valued functions even the comparison of ground terms can be
undecidable (see Example~\ref{ex:realKBO} in Section \ref{sec:main:result}).


The main result of this \paper\ is the following.

  \begin{center}\fbox{%
  \begin{minipage}{0.7\textwidth}
    \noindent
    \textbf{Theorem~\ref{thm:main}:}
    \emph{The existential first-order theory of any term algebra with the
    Knuth-Bendix order in a signature with at least two symbols is
    NP-complete.}
  \end{minipage}}
  \end{center}

To prove this result, we introduce a notion of Knuth-Bendix ordering
constraint and show the following.


  \begin{center}\fbox{%
  \begin{minipage}{0.7\textwidth}
    \noindent
    \textbf{Theorem~\ref{thm:KBOisNP}:}
 \emph{For every Knuth-Bendix order, the problem of solving ordering constraints
  is contained in NP.}
  \end{minipage}}
  \end{center}

We also show that the systems of linear Diophantine equations and inequalities
can be represented as ordering constraints for some Knuth--Bendix orders, 
and as a corollary we obtain the following.

  \begin{center}\fbox{%
  \begin{minipage}{0.7\textwidth}
    \noindent
    \textbf{Theorem~\ref{thm:KBOisNPcomp}:}
    \emph{
     For some Knuth-Bendix orders, the problem of solving ordering
    constraints is NP-complete.}
  \end{minipage}}
  \end{center}

The proof of Theorem~\ref{thm:main} will be given after a series of lemmas. The idea of the proof is as
follows. First, we will make $\TA(\Sigma)$ into a two-sorted structure by
adding the sort of natural numbers, and extend its signature by

\begin{enumerate}
  \item the weight function $|\cdot|$ on ground terms;

  \item the addition function $+$ on natural numbers;

  \item the Knuth-Bendix order $\KBo$ on ground terms.
\end{enumerate}
Given an existential formula of the first-order theory of a term algebra with
the Knuth-Bendix order, we will transform it step by step into an equivalent
disjunction of existential formulas of the extended signature. The main aim
of these steps is to replace all occurrences of $\KBo$ by linear Diophantine
inequalities on the weights of variables. After such a transformation we will
obtain existential formulas consisting of linear Diophantine inequalities on
the weight of variables plus statements expressing that, for some fixed
natural number $N$, there exists at least $N$ terms of the same weight as
$|x|$, where $x$ is a variable. We will show how these statements can be
expressed using systems of linear Diophantine inequalities on the weights of
variables and then use the fact that the decidability of systems of linear
Diophantine equations is in NP.

We denote by \DI{$\TA^+(\Sigma)$}{TA@$\protect\TA^+(\Sigma)$} the following
structure with two sorts: the \DI{term algebra sort}{term algebra sort} and
the \DI{arithmetical sort}{arithmetical sort}. The domains of the term
algebra sort and the arithmetical sort are the sets of ground terms of
$\Sigma$ and natural numbers, respectively. The signature of $\TA^+(\Sigma)$
consists of

\begin{enumerate}
  \item all symbols of $\Sigma$ interpreted as in $\TA(\Sigma)$;

  \item symbols $0,1,>,+$ having their conventional interpretation over
  natural numbers;

  \item the binary relation symbol $\KBo$ on the term algebra sort,
  interpreted as the Knuth-Bendix order;

  \item the unary function symbol $|\cdot|$, interpreted as the weight
  function mapping terms to numbers.
\end{enumerate}
When we need to distinguish the equality $=$ on the term algebra sort from
the equality on the arithmetical sort, we denote the former by
\DI{$=_{\TA}$}{$=_{\protect\TA}$}, and the latter by
\DI{$=_{\nat}$}{$=_{\protect\nat}$}.

We will prove that the existential theory of $\TA^+(\Sigma)$ is in NP, from
which the fact that the existential theory of any term algebra with the
Knuth-Bendix order belongs to NP follows immediately. We consider
\DI{satisfiability}{satisfiability}, \DI{validity}{validity}, and
\DI{equivalence}{equivalence} of formulas with respect to the structure
$\TA^+(\Sigma)$. We call a \DI{constraint}{constraint} in the language of
$\TA^+(\Sigma)$ any conjunction of atomic formulas of this language.

\begin{Lemma}
\label{lem:E-theory=satisfiability}%
  The existential theory of $\TA^+(\Sigma)$ is in NP if and only if so is
  the constraint satisfiability problem.
\end{Lemma}
\begin{proof}
  Obviously any instance $A$ of the constraint satisfiability problem can be
  considered as validity of the existential sentence $\exists x_1 \ldots x_n
  A$, where $\xone{x}{n}$ are all variables of $A$, so the ``only if''
  direction is trivial.

  To prove the ``if'' direction, take any existential formula $\exists
  \xone{x}{n} A$. This formula is satisfiable if and only if so is the
  quantifier-free formula $A$. By converting $A$ into disjunctive normal form
  we can assume that $A$ is built from literals using $\andl,\orl$. Replace
  in $A$

  \begin{enumerate}
    \item any formula $\notl s \KBo t$ by $s =_{\TA} t \orl t \KBo s$,

    \item any formula $\notl s =_{\TA} t$ by $s \KBo t \orl t \KBo s$,

    \item any formula $\notl p > q$ by $p=_{\nat}q \orl q > p$,

    \item any formula $\notl p =_{\nat} q$ by $p > q \orl q > p$,
  \end{enumerate}
  and convert $A$ into disjunctive normal form again. It is easy to see that
  we obtain a disjunction of constraints. The transformation gives an
  equivalent formula since both orders $\KBo$ and $>$ are total.

  It follows from these arguments that there exists a nondeterministic
  polynomial-time algorithm which, given an existential sentence $A$,
  computes on every branch a constraint $C_i$ such that $A$ is valid if and
  only if one of the constraints $C_i$ is satisfiable.
\end{proof}

A \DI{substitution}{substitution} is a mapping from the set of variables to
the set of terms. A substitution $\theta$ is called \DII{grounding}{grounding
substitution}{substitution!grounding} for an expression $C$ (i.e., term or
constraint) if for every variable $x$ occurring in $C$ the term $\theta(x)$
is ground. Let $\theta$ be a substitution grounding for an expression $C$. We
denote by $C\theta$ the expression obtained from $C$ by replacing in it every
variable $x$ by $\theta(x)$. A substitution $\theta$ is called a
\DI{solution}{solution} to a constraint $C$ if $\theta$ is grounding for $C$
and $C\theta$ is valid in $\TA^+(\Sigma)$.

In the sequel we will often replace a constraint $C(\bar{x})$ by a formula
$A(\bar{x},\bar{y})$ containing extra variables $\bar{y}$ and say that they
are ``equivalent''. By this we mean that $\TA^+(\Sigma) \models \forall
\bar{x} (C(\bar{x}) \iffl \exists \bar{y} A(\bar{x},\bar{y}))$. In other
words, the set of solutions to $C$ is exactly the set solutions to $A$
projected on $\bar{x}$.

                          \section{Isolated forms}
                          \label{sec:isolated:forms}

We are interested not only in satisfiability of constraints, but also in
their solutions. Our algorithm will consist of equivalence-preserving
transformation steps. When the signature contains no unary function symbol of
the weight $0$, the transformation will preserve equivalence in the following
strong sense. At each step, given a constraint $C(\bar{x})$, we transform it
into constraints $C_1(\bar{x},\bar{y}),\ldots,C_n(\bar{x},\bar{y})$ such that
for every sequence of ground terms $\bar{t}$, the constraint $C(\bar{t})$
holds if and only if there exist $k$ and a sequence of ground terms $\bar{s}$
such that $C_k(\bar{t},\bar{s})$ holds. In other words, the following formula
holds in $\TA^+(\Sigma)$:

  \[
    C(\bar{x}) \iffl
      \exists \bar{y}
        ( C_1(\bar{x},\bar{y}) \orl \ldots \orl C_n(\bar{x},\bar{y}) ).
  \]
Moreover this transformations will be presented as a nondeterministic
polynomial-time algorithm which computes on every branch some
$C_i(\bar{x},\bar{y})$, and every $C_i(\bar{x},\bar{y})$ is computed on at
least one branch. When the signature contains a unary function symbol of the
weight $0$, the transformation will preserve a weaker form of equivalence:
some solutions will be lost, but solvability will be preserved. More
precisely, we will introduce a notion of an $f$-variant of a term and show
that the following formula holds:

  \begin{equation}
  \label{eq:f-equiv}
    C(\bar{x}) \iffl
      \exists \bar{y} \exists \bar{z}
        ( \textit{f-variant}(\bar{x},\bar{z}) \andl
          (C_1(\bar{z},\bar{y}) \orl \ldots \orl C_n(\bar{z},\bar{y}) ) ),
  \end{equation}
where $\textit{f-variant}(\bar{x},\bar{z})$ expresses that $\bar{x}$ and
$\bar{z}$ are $f$-variants.

In our proof, we will reduce solvability of Knuth-Bendix ordering constraints
to the problem of solvability of systems of linear Diophantine inequalities
on the weights of variables. Condition \ref{cond:KB:1} of the definition of
the Knuth-Bendix order $|t| > |s|$ has a simple translation into a linear
Diophantine inequality, but conditions \ref{cond:KB:2} and \ref{cond:KB:3} do
not have. So we will split the Knuth-Bendix order in two partial orders:
$\KBw$ corresponding to condition \ref{cond:KB:1} and $\KBe$ corresponding to
conditions \ref{cond:KB:2} and \ref{cond:KB:3}. Formally, we denote by \DI{$t
\KBw s$}{$\protect\KBw$} the formula $|t|>|s|$ and by \DI{$t \KBe
s$}{$\protect\KBe$} the formula $|t|=_{\nat}|s| \andl t \KBo s$. Obviously,
$t_1 \KBo t_2$ if and only if $ t_1 \KBe t_2 \vee t_1 \KBw t_2$. So in the
sequel we will assume that $\KBo$ is replaced by the new symbols $\KBe$ and
$\KBw$.

We use $x_1 \KBo x_2 \KBo \ldots \KBo x_n$ to denote the formula $x_1 \KBo
x_2 \wedge x_2 \KBo x_3 \wedge \ldots \wedge  x_{n-1} \KBo x_n$, and similar
for other binary symbols in place of $\KBo$.

A term $t$ is called \DI{flat}{flat term} if $t$ is either a variable or has
the form $g(\xone{x}{m})$, where $g \in \Sigma$, $m \geq 0$, and
$\xone{x}{m}$ are variables. We call a constraint \DI{chained}{chained
constraint} if

\begin{enumerate}
  \item it has a form $t_1 \# t_2 \# \ldots \# t_n$, where each occurrence of
  $\#$ is $\KBw$, $\KBe$ or $=_{\TA}$;

  \item each term $t_i$ is flat;

  \item if some of the $t_i$'s has the form $g(\xone{x}{n})$, then
  $\xone{x}{n}$ are some of the $t_j$'s.
\end{enumerate}
Denote by \DI{$\false$}{$\protect\false$} the logical constant ``false''.

\begin{Lemma}
\label{lem:chained}%
  Any constraint $C$ is equivalent to a disjunction $C_1 \orl \ldots \orl
  C_k$ of chained constraints. Moreover, there exists a nondeterministic
  polynomial-time algorithm which, for a given $C$, computes on every branch
  either $\false$ or some $C_i$; and every $C_i$ is computed on at least one
  branch.
\end{Lemma}
\begin{proof}
  First, we can apply flattening to all terms occurring in $C$ as follows. If
  a nonflat term $g(\xone{t}{m})$ occurs in $C$, take any $i$ such that $t_i$
  is not a variable. Then replace $C$ by $v = t_i \andl C'$, where $v$ is a
  new variable and $C'$ is obtained from $C$ by replacing all occurrences of
  $t_i$ by $v$. After a finite number of such replacements all terms will
  become flat.

  Let $s,t$ be flat terms occurring in $C$ such that no comparison $s \# t$
  occurs in $C$. Using the valid formula $s \KBw t \orl s \KBe t \orl s
  =_{\TA} t \orl t \KBw s \orl t \KBe s$ we can replace $C$ by the
  disjunction of the constraints

    \[
      \begin{array}{lll}
        s \KBw t \andl C, &
        s \KBe t \andl C, &
        s =_{\TA} t \andl C, \\
        t \KBw s \andl C, &
        t \KBe s \andl C.
      \end{array}
    \]
  By repeatedly doing this transformation we obtain a disjunction of
  constraints $C_1 \orl \ldots \orl C_k$ in which for every terms $s,t$ and
  every $i \in \setof{1,\ldots,k}$ some comparison constraint $s \# t$ occurs
  in $C_i$.

  To complete the proof we show how to turn each $C_i$ into a chained
  constraint. Let us call a \emph{cycle} any constraint $s_1 \# s_2 \# \ldots
  \# s_n \# s_1$, where $n \geq 1$. We can remove all cycles from $C_i$ using
  the following observation:

  \begin{enumerate}
    \item if all $\#$ in the cycle are $=_{\TA}$, then $s_n \# s_1$ can be
    removed from the constraint;

    \item if some $\#$ in the cycle is $\KBw$ or $\KBe$, then the constraint
    $C_i$ is unsatisfiable.
  \end{enumerate}
  After removal of all cycles the constraint $C_i$ can still be not chained
  because it can contain \emph{transitive subconstraints} of the form $s_1 \#
  s_2 \# \ldots \# s_n \andl s_1 \# s_n$, $n \geq 2$. Then either $C_i$ is
  unsatisfiable or $s_1 \# s_n$ can be removed using the following
  observations:

  \begin{enumerate}
    \item \emph{Case: $s_1 \# s_n$ is $s_1 \KBw s_n$.} If some $\#$ in $s_1
    \# s_2 \# \ldots \# s_n$ is $\KBw$, then $s_1 \KBw s_n$ follows from $s_1
    \# s_2 \# \ldots \# s_n$, otherwise $s_1 \# s_2 \# \ldots \# s_n$ implies
    $|s_1|=|s_n|$ and hence $C_i$ is unsatisfiable.

    \item \emph{Case: $s_1 \# s_n$ is $s_1 \KBe s_n$.} If some $\#$ in $s_1
    \# s_2 \# \ldots \# s_n$ is $\KBw$, then $C_i$ is unsatisfiable. If all
    $\#$ in $s_1 \# s_2 \# \ldots \# s_n$ are $=_{\TA}$, then $C_i$ is
    unsatisfiable too. Otherwise, all $\#$ in $s_1 \# s_2 \# \ldots \# s_n$
    are either $\KBe$ or $=_{\TA}$, and at least one of them is $\KBe$. It is
    not hard to argue that $s_1 \KBe s_n$ follows from $s_1 \# s_2 \# \ldots
    \# s_n$.

    \item \emph{Case: $s_1 \# s_n$ is $s_1 =_{\TA} s_n$.} If all $\#$ in $s_1
    \# s_2 \# \ldots \# s_n$ are $=_{\TA}$, then $s_1 =_{\TA} s_n$ follows
    from $s_1 \# s_2 \# \ldots \# s_n$, otherwise $C_i$ is unsatisfiable.
  \end{enumerate}
  It is easy to see that after the removal of all cycles and transitive
  subconstraints the constraint $C_i$ becomes chained.

  Note that the transformation of $C$ into the disjunction of constraints
  $C_1 \orl \ldots \orl C_k$ in the proof can be done in nondeterministic
  polynomial time in the following sense: there exists a nondeterministic
  polynomial-time algorithm which, given $C$ computes on every branch either
  $\false$ or some $C_i$, and every $C_i$ is computed on at least one branch.
\end{proof}
We will now introduce several special kinds of constraints which will be used
in our proofs below, namely \emph{arithmetical}, \emph{triangle},
\emph{simple}, and \emph{isolated}.

A constraint is called \emph{arithmetical} if it uses only arithmetical
relations $=_{\nat}$ and $>$, for example $|f(x)| > |a|+3$.

A constraint $y_1 =_{\TA} t_1 \andl \ldots \andl y_n =_{\TA} t_n$ is said to
be in \DI{triangle form}{triangle form} if

\begin{enumerate}
  \item $\xone{y}{n}$ are pairwise different variables, and

  \item for all $j \geq i$ the variable $y_i$ does not occur in $t_j$.
\end{enumerate}
The variables $\xone{y}{n}$ are said to be \DI{dependent}{dependent} in this
constraint.

A constraint is said to be \DI{simple}{simple} if it has the form

  \[
    x_{11} \KBe x_{12} \KBe \ldots \KBe x_{1n_1}
      \andl \ldots \andl
    x_{k1} \KBe x_{k2} \KBe \ldots \KBe x_{kn_k},
  \]
where $x_{11},\ldots,x_{kn_k}$ are pairwise different variables.

A constraint is said to be in \DI{isolated form}{isolated form} if either it
is $\false$ or it has the form

  \[
    \Arith \andl \Triang \andl \Simp,
  \]
where $\Arith$ is an arithmetical constraint, $\Triang$ is in triangle
form, and $\Simp$ is a simple constraint such that no variable of $\Simp$ is
dependent in $\Triang$.

Our decision procedure for the Knuth-Bendix ordering constraints is designed
as follows. By Lemma~\ref{lem:chained} we can transform any constraint into
an equivalent disjunction of chained constraints. Our next step is to give a
transformation of any chained constraint into an equivalent disjunction of
constraints in isolated form. Then in Section~\ref{sec:isolated->diophantine}
we show how to transform any constraint in isolated form into an equivalent
disjunction of systems of linear Diophantine inequalities on the weights of
variables. Then we can use the result that the decidability of systems of
linear Diophantine inequalities is in NP.

Let us show how to transform any chained constraint into an equivalent
disjunction of isolated forms. The transformation will work on the
constraints of the form

  \begin{equation}
    \label{eqn:chained}%
    \Cha \andl \Arith \andl \Triang \andl \Simp,
  \end{equation}
such that

\begin{enumerate}
  \item $\Arith,\Triang,\Simp$ are as in the definition of isolated form;

  \item $\Cha$ is a chained constraint;

  \item each variable of $\Cha$ neither occurs in $\Simp$ nor is dependent in
  $\Triang$.
\end{enumerate}
We will call such constraints \eqref{eqn:chained} \DI{working}{working
constraint}. Let us call the \DI{size}{size} of a chained constraint $C$ the
total number of occurrences of function symbols and variables in $C$.
Likewise, the \DI{essential size}{essential size} of a working constraint is
the size of its chained part $\Cha$.

At each transformation step we will replace working constraint
\eqref{eqn:chained} by a disjunction of working constraints but of smaller
essential sizes. Evidently, when the essential size is $0$, we obtain a
constraint in isolated form.

Let us prove some lemmas about solutions to constraints of the form
\eqref{eqn:chained}. Note that any chained constraint is of the form

  \begin{equation} \label{eqn:chained1}
    \begin{array}{c}
      t_{11} \# t_{12} \#  \ldots \# t_{1m_1}    \\
                      \KBw                     \\
                      \cdots                   \\
                      \KBw                     \\
      t_{k1} \# t_{k2} \# \ldots \# t_{km_k},
    \end{array}
  \end{equation}
where each $\#$ is either $=_{\TA}$ or $\KBe$ and each $t_{ij}$ is a flat
term. We call a \DI{row}{row} in such a constraint any maximal subsequence
$t_{i1} \# t_{i2} \# \ldots \# t_{im_i}$ in which $\KBw$ does not occur. So
constraint \eqref{eqn:chained1} contains $k$ rows, the first one is $t_{11}
\# t_{12} \#  \ldots \# t_{1m_1}$ and the last one $t_{k1} \# t_{k2} \#
\ldots \# t_{km_k}$. Note that for any solution to \eqref{eqn:chained1} all
terms in a row have the same weight.

\begin{Lemma}
\label{lem:one:occurrence}%
  There exists a polynomial-time algorithm which transforms any chained
  constraint $C$ into an equivalent chained constraint $C'$ such that the
  size of $C'$ is not greater than the size of $C$, either $C'$ is $\false$
  or of the form \eqref{eqn:chained1}, and $C'$ has the following property.
  Suppose some term of the first row $t_{1j}$ of $C'$ is a variable $y$. Then
  either

  \begin{enumerate}
    \item $y$ has exactly one occurrence in $C'$, namely $t_{1j}$ itself; or

    \item $y$ has exactly two occurrences in $C'$, both in the first row:
    some $t_{1n}$ has the form $f(y)$ for $n < j$, and $w(f) = 0$; moreover
    in this case there exists at least one $\KBe$ between $t_{1n}$ and
    $t_{1j}$.
  \end{enumerate}
\end{Lemma}
\begin{proof}
  Note that if $y$ occurs in any term $t(y)$ which is not in the first row,
  then $C$ is unsatisfiable, since for any solution $\theta$ to $C$ we have
  $|y\theta| > |t(y)\theta|$, which is impossible. Suppose that $y$ has
  another occurrence in a term $t_{1n}$ of the first row. Consider two cases.

  \begin{enumerate}
    \item \emph{$t_{1n}$ coincides with $y$}. Then either $C$ has no
    solution, or part of the first row between $t_{1n}$ and $t_{1j}$ has the
    form $y =_{\TA} \ldots =_{\TA} y$. In the latter case part $y = _{\TA}$
    can be removed from the first row, so we can assume that no term in the
    first row except $t_{1j}$ is $y$.

    \item \emph{$t_{1n}$ is a nonvariable term containing $y$.} Since
    $t_{1n}$ and $y$ are in the same row, for every solution $\theta$ to $C$
    we have $|y\theta| = |t_{1n}\theta|$. Since $t_{1n}$ is a flat term, by
    Lemma~\ref{lem:weight:props} the equality $|y\theta| = |t_{1n}\theta|$ is
    possible only if $t_{1n}$ is $f(y)$ and $n < j$. Finally, if $f(y)$ has
    more than one occurrences in the first row, we can get rid of all of them
    but one in the same way as we got rid of multiple occurrences of $y$.
  \end{enumerate}
  Note that the transformation presented in this proof can be made in
  polynomial time. It is also not hard to argue that the transformation does
  not increase the size of the constraint.
\end{proof}

We will now take a working constraint $\Cha \andl \Arith \andl \Triang \andl
\Simp$, whose chained part satisfies Lemma~\ref{lem:one:occurrence} and
transform it into an equivalent disjunction of working constraints of smaller
essential sizes in Lemma~\ref{lem:into:isolated} below. More precisely, these
constraints will be equivalent when the signature contains no unary function
symbol of the weight $0$. When the signature contains such a symbol $f$, a
weaker notion of equivalence will hold, see formula \eqref{eq:f-equiv} on
page \pageref{eq:f-equiv}.

A term $s$ is called an \DI{$f$-variant}{f-variant@$f$-variant} of a term $t$
if $s$ can be obtained from $t$ by a sequence of operations of the following
forms: replacement of a subterm $f(r)$ by $r$ or replacement of a subterm $r$
by $f(r)$. Evidently, $f$-variant is an equivalence relation. Two substitutions
$\theta_1$ and $\theta_2$ are said to be $f$-variants if for every variable
$x$ the term $x\theta_1$ is an $f$-variant of $x\theta_2$. In the proof of
several lemmas below we will replace a constraint $C(\bar{x})$ by a formula
$A(\bar{x},\bar{y})$ containing extra variables $\bar{y}$ and say that
$C(\bar{x})$ and $A(\bar{x},\bar{y})$ are \DI{equivalent up to
$f$}{equivalent up to $f$}. By this we mean the following.

\begin{enumerate}
  \item For every substitution $\theta_1$ grounding for $\bar{x}$ such that
  $\TA^+(\Sigma) \models C(\bar{x})\theta_1$, there exists a substitution
  $\theta_2$ grounding for $\bar{x},\bar{y}$ such that $\TA^+(\Sigma) \models
  A(\bar{x},\bar{y})\theta_2$, and the restriction of $\theta_2$ to $\bar{x}$
  is an $f$-variant of $\theta_1$.

  \item For every substitution $\theta_2$  grounding
  for $\bar{x},\bar{y}$ such that
  $\TA^+(\Sigma) \models A(\bar{x},\bar{y})\theta_2$,
  there exists a substitution $\theta_1$ such that $\TA^+(\Sigma) \models
  C(\bar{x})\theta_1$ and $\theta_1$ is an $f$-variant of the restriction of
  $\theta_2$ to $\bar{x}$. In other words, formula \eqref{eq:f-equiv} on page
  \pageref{eq:f-equiv} holds.
\end{enumerate}
Note that when the signature contains no unary function symbol of the weight
$0$, equivalence up to $f$ is the same as equality of terms in
$\TA^+(\Sigma)$.

\begin{Lemma}
  Let $C = \Cha \andl \Arith \andl \Triang \andl \Simp$ be a working
  constraint and $\theta_1$ be a solution to $C$. Let $\theta_2$ be an
  $f$-variant of $\theta_1$ such that

  \begin{enumerate}
    \item $\theta_2$ is a solution to $\Cha$ and
    \item $\theta_2$ coincides with $\theta_1$ on all variables not occurring
    in $\Cha$.
  \end{enumerate}
  Then there exists an $f$-variant $\theta_3$ of
  $\theta_2$ such that

  \begin{enumerate}
    \item $\theta_3$ is a solution to $C$ and

    \item $\theta_3$ coincides with $\theta_2$ on all variables except for
    the dependent variables of $\Triang$.
  \end{enumerate}
\end{Lemma}
\begin{proof}
  Let us first prove that $\theta_2$ is a solution to both $\Arith$ and
  $\Simp$. Since $\Simp$ and $\Cha$ have no common variables, it follows that
  $\theta_1$ and $\theta_2$ agree on all variables of $\Simp$, and so
  $\theta_2$ is a solution to $\Simp$. Since $\theta_1$ and $\theta_2$ are
  $f$-variants and the weight of $f$ is $0$, for every term $t$ we have
  $|t\theta_1| = |t\theta_2|$, whenever $t\theta_1$ is ground. Therefore,
  $\theta_2$ is a solution to $\Arith$ if and only if so is $\theta_1$. So
  $\theta_2$ is a solution to $\Arith$.

  It is fairly easy to see that $\theta_2$ can be changed on the dependent
  variables of $\Triang$ obtaining a solution $\theta_3$ to $C$ which
  satisfies the conditions of the lemma.
\end{proof}
This lemma will be used below in the following way. Instead of considering
the set $\Theta_1$ of all solutions to $\Cha$ we can restrict ourselves to a
subset $\Theta_2$ of $\Theta_1$ as soon as for every solution $\theta_1 \in
\Theta_1$ there exists a solution $\theta_2 \in \Theta_2$ such that
$\theta_2$ is an $f$-variant of $\theta_1$.

Let us call an \DI{$f$-term}{f-term@$f$-term} any term of the form $f(t)$. By
the \DI{$f$-height}{f-height@$f$-height} of a term $t$ we mean the number $n$
such that $t = f^n(s)$ and $s$ is not an $f$-term. Note that the $f$-terms
are exactly the terms of a positive $f$-height. We call the
\DI{$f$-distance}{f-distance@$f$-difference} between two terms $s$ and $t$
the difference between the $f$-height of $s$ and $f$-height of $t$. For
example, the $f$-distance between the terms $f(a)$ and $f(f(g(a,b))$ is $-1$.

Let us now prove a lemma which implies that any solution to $C$ can be
transformed into a solution with a ``small'' $f$-height.

\begin{Lemma}
\label{lem:f-height}%
  Let $\Cha$ be a chained constraint of the form

    \[
        p_{l} \# p_{l-1} \#  \ldots \# p_1  \KBw \ldots ,
    \]
  where each $\#$ is either $=_{\TA}$ or $\KBe$. Further, let $\Cha$ satisfy
  the conditions of Lemma~\ref{lem:one:occurrence} and $\theta$ be a solution
  to $\Cha$. Then there exists an $f$-variant $\theta'$ of $\theta$ such that

  \begin{enumerate}
    \item $\theta'$ is a solution to $\Cha$ and

    \item for every $k \in \setof{1,\ldots,l}$, the $f$-height of
    $p_k\theta'$ is at most $k$.
  \end{enumerate}
\end{Lemma}
\begin{proof}
  Let us first prove the following statement

  \begin{eqitemize}
    \item The row $p_l \# p_{l-1} \# \ldots \# p_1$ has a solution
    $\theta_1$, such that (i) $\theta_1$ is an $f$-variant of $\theta$, (ii)
    for every $1 < k \leq l$ the $f$-distance between $p_k\theta_1$ and
    $p_{k-1}\theta_1$ is at most $1$. \label{itm:z1}
  \end{eqitemize}
  Suppose that for some $k$ the $f$-distance between $p_k\theta$ and
  $p_{k-1}\theta$ is $d > 1$. Evidently, to prove \eqref{itm:z1} it is
  enough to show the following.

  \begin{eqitemize}
    \item There exists a solution $\theta_2$ such that (i) $\theta_2$ is an
    $f$-variant of $\theta$, (ii) the $f$-distance between $p_k\theta_2$ and
    $p_{k-1}\theta_2$ is $d - 1$, and (iii) for every $k' \neq k$ the
    $f$-distance between $p_{k'}\theta_2$ and $p_{k'-1}\theta_2$ coincides
    with the $f$-distance between $p_{k'}\theta$ and $p_{k'-1}\theta$.
    \label{itm:z2}
  \end{eqitemize}
  Let us show \eqref{itm:z2}, and hence \eqref{itm:z1}. Since $\theta$ is a
  solution to the row, then for every $k''' \geq k$ the $f$-distance between
  any $p_{k'''}\theta$ and $p_k\theta$ is nonnegative. Likewise, for every
  $k'' < k-1$ the $f$-distance between any $p_{k-1}\theta$ and
  $p_{k''}\theta$ is nonnegative. Therefore, for all $k''' \geq k > k''$, the
  $f$-distance between $p_{k'''}\theta$ and $p_{k''}\theta$ is $\geq d$, and
  hence is at least $2$. Let us prove the following.

  \begin{eqitemize}
    \item Every variable $x$ occurring in $p_l \# p_{l-1} \# \ldots \# p_k$
    does not occur in $p_{k-1} \# \ldots \# p_1$. \label{itm:z3}
  \end{eqitemize}
  Let $x$ occur in terms $p_i$ and $p_j$ such that $l \geq i \geq k$ and $k-1
  \geq j \geq 1$. Since the constraint satisfies
  Lemma~\ref{lem:one:occurrence}, then $p_i = f(x)$ and $p_j=x$. Then the
  $f$-distance between $p_i\theta$ and $p_j\theta$ is $1$, but by our
  assumption it is at least $2$, so we obtain a contradiction. Hence
  \eqref{itm:z3} is proved.

  Now note the following.

  \begin{eqitemize}
    \item If for some $k''' \geq k$ a variable $x$ occurs in $p_{k'''}$, then
    $x\theta$ is an $f$-term. \label{itm:z4}
  \end{eqitemize}
  Suppose, by contradiction, that $x\theta$ is not an $f$-term. Note that
  $p_{k'''}\theta$ has a positive $f$-height, so $p_{k'''}$ is either $x$ of
  $f(x)$. But we proved before that the $f$-distance between $p_{k'''}$ and
  $p_{k-1}$ is at least $2$, so $x$ must be an $f$-term.

  Now, to satisfy \eqref{itm:z2}, define the substitution $\theta_2$ as
  follows:

    \[
      \theta_2(x) =
        \left\{
          \begin{array}{ll}
            \theta(x), &
              \begin{array}{l}
              \text{if $x$ does not occur in $p_l,\ldots,p_k$};
              \end{array} \\
            t, &
              \begin{array}[t]{l}
                \text{if $x$ occurs in $p_l,\ldots,p_k$} 
                \text{ and } \theta(x) = f(t).
              \end{array}
          \end{array}
        \right.
    \]
  By \eqref{itm:z3} and \eqref{itm:z4}, $\theta_2$ is defined correctly. We
  claim that $\theta_2$ satisfies \eqref{itm:z2}. The properties (i)-(iii) of
  \eqref{itm:z2} are straightforward by our construction, it only remains to
  prove that $\theta_2$ is a solution to the row, i.e.\ for every $k'$ we
  have $p_{k'}\theta_2 \# p_{k'-1}\theta_2$. Well, for $k' > k$ we have
  $p_{k'}\theta = f(p_{k'}\theta_2)$ and $p_{k'-1}\theta =
  f(p_{k'-1}\theta_2)$, and for $k' < k$ we have $p_{k'}\theta =
  p_{k'}\theta_2$ and $p_{k'-1}\theta = p_{k'-1}\theta_2$, in both cases
  $p_{k'}\theta_2 \# p_{k'-1}\theta_2$ follows from $p_{k'}\theta \#
  p_{k'-1}\theta$. The only difficult case is $k=k'$.

  Assume $k = k'$. Since the $f$-distance between $p_k\theta$ and
  $p_{k-1}\theta$ is $d > 1$, we have $p_k\theta \neq p_{k-1}\theta$, and
  hence $p_k \# p_{k-1}$ must be $p_k \KBe p_{k-1}$. Since $\theta$ is a
  solution to $p_k \KBe p_{k-1}$ and since $\theta_2$ is an $f$-variant of
  $\theta$, the weights of $p_k\theta_2$ and $p_{k-1}\theta_2$ coincide. But
  then $p_k\theta_2 \KBe p_{k-1}\theta_2$ follows from the fact that the
  $f$-distance between $p_k\theta_2$ and $p_{k-1}\theta_2$ is $d-1 \geq 1$.

  Now the proof of \eqref{itm:z2}, and hence of \eqref{itm:z1}, is
  completed. In the same way as \eqref{itm:z1}, we can also prove

  \begin{eqitemize}
    \item The constraint $\Cha$ has a solution $\theta'$ such that (i) $\theta'$
    is an $f$-variant of $\theta$, (ii) for every $1 < k \leq l$ the
    $f$-distance between $p_k\theta_1$ and $p_{k-1}\theta'$ is at most $1$.
    (iii) the $f$-height of $p_1\theta'$ is at most $1$; (iv) $\theta'$ and
    $\theta$ coincide on all variables occurring in the rows below the first
    one. \label{itm:z5}
  \end{eqitemize}
  It is not hard to derive Lemma~\ref{lem:f-height} from \eqref{itm:z5}.
\end{proof}

The following lemma is the main (and the last) lemma of this section.

\begin{Lemma}
\label{lem:into:isolated}%
  Let $C = \Cha \andl \Arith \andl \Triang \andl \Simp$ be a working
  constraint in which $\Cha$ is nonempty. There exists a nondeterministic
  polynomial-time algorithm which transforms $C$ into a disjunction of
  working constraints having $\Cha$ of smaller sizes and equivalent to $C$ up
  to $f$.
\end{Lemma}
\begin{proof}
  The proof is rather complex, so we will give a plan of it. The proof is
  presented as a series of transformations on the first row of $\Cha$. These
  transformations may result in new constraints added to $\Arith$, $\Triang$,
  and $\Simp$. First, we will get rid of equations $s =_{\TA} t$ in the first
  row, by introducing \emph{quasi-flat} terms, i.e.\ terms $f^k(t)$, where
  $t$ is flat. If the first row contained no function symbols, then we will
  replace the first row by new constraints added to $\Simp$ and $\Arith$,
  thus decreasing the size of the chained part. If there were function
  symbols in the first row, we will continue as follows.

  We will ``guess'' the values of some variables $x$ of the first row, i.e.\
  replace them by some quasi-flat term $f^m(g(\bar{y}))$, where $\bar{y}$ is
  a sequence of new variables. After these steps, the size of the first row
  can, in general, increase. Then we will show how to replace the first row
  by new constraints involving only variables occurring in the row, but not
  function symbols. Finally, we will prove that the number of variables from
  the new constraints that remain in the chained part is not greater than the
  original number of variables in the first row, and therefore the size of
  the chained part decreases.

  Formally,
  consider the first row of $\Cha$. Let this row be $p_l \# p_{l-1} \# \ldots
  \# p_1$. Then  $\Cha$ has the form $p_l \# p_{l-1} \# \ldots \# p_1 \KBw
  t_1 \# \ldots \# t_n$. If $l=1$, i.e., the first row consists of one term,
  we can remove this row and add $|p_1| > |t_1|$ to $\Arith$ obtaining an
  equivalent constraint with smaller essential size, that is, the size of
  $\Cha$. So we assume that the first row contains at least two terms.

  As before, we assume that $f$ is a unary function symbol of the weight $0$.
  By Lemma~\ref{lem:f-height}, if some $p_i$ is either a variable $x$ or a
  term $f(x)$, it is enough to search for solutions $\theta$ such that the
  height of $x\theta$ is at most $l$.

  A term is called \DI{quasi-flat}{quasi-flat term} if it has the form
  $f^k(t)$ where $t$ is flat. We will now get rid of equalities in the first
  row, but by introducing quasi-flat terms instead of the flat ones. When we
  use notation $f^k(t)$ below, we assume $k \geq 0$, and $f^0(t)$ will stand
  for $t$. 
%
We eliminate equalities from  the first
  row  in two steps. First we will eliminate equalities among 
  variables and $f$--terms transforming them into an equivalent set of 
  equalities in triangle form, 
  then we eliminate all other equalities in the first row.

   Consider the set $S$ of all equalities $t=_{\TA}s$ occurring in the
   first row of $\Cha$, where $s$ and $t$ are either variables or flat
   $f$--terms. 
   We will transform $S$ into an equivalent system $F$ in triangle      
   form such that all terms in $F$ will be flat. We assume that before
  the transformation $F$ is empty. 
   First we replace all equalities in $S$ of the form $f(x)=_{\TA}f(y)$
   by $x=_{\TA}y$ obtaining an equivalent system $S'$ 
   in which all equalities are of the form $x=_{\TA}t$. 
   Now, either $S'$ is unsatisfiable or there exists 
   an equality $x=_{\TA}t$ in $S'$, such that 
   $x$ does not occur in $f$--terms of $S'$. 
   We move such an equality  $x=_{\TA}t $ 
   into $F$ and replace all occurrences
   of $x$ in $S'$ by $t$, obtaining $S''$. It is easy to see  that 
   the system $F\cup S''$ is equivalent to $S$, all terms in $F\cup S''$ 
   are flat, $F$ is in triangle form  
   and the number of variables occurring into $S''$ is less than
   the number of variables occurring into $S$. 
   Repeating this process 
   we can eliminate all  variables from $S$ and obtain the required
  $F$ in polynomial time.

 Now we remove from $\Cha$ all equalities occurring in $S$.
 Let us note that variables of $F$ can occur in $\Cha$  
  only in  the first row, and only in the terms $f^r(y)$ 
  for $0\leq r\leq 1$.  
  Next we repeatedly replace all occurrences of dependent variables of $F$
  occurring in $\Cha$ obtaining an equivalent constraint in chained form 
  with terms of the form $f^k(x)$ where $k$ is bounded by the size 
  of $F$. Finally we move $F$ into $\Triang$.

  After all these transformations we can assume that equalities $f^k(x)
  =_{\TA} f^m(y)$ do not occur in the first row.

  If the first row contains an equality $x =_{\TA} t$ between a variable and
  a term, we replace this equality by $t$, replace all occurrences of $x$ by
  $t$ in the first row, and add $x=_{\TA}t$ to $\Triang$ obtaining an
  equivalent working constraint.
  Since $x$ can occur only in the terms of the form  $f^{r}(x)$, it is
  easy to see that these replacements can be done in polynomial time. 

  If the first row contains an equality $g(\xone{x}{m}) =_{\TA}
  h(\xone{t}{n})$ where $g$ and $h$ are different function symbols, the
  constraint is unsatisfiable.

  If the first row contains an equality $g(\xone{x}{n}) =_{\TA}
  g(\xone{y}{n})$ we do the following. If the term $g(\xone{x}{n})$ coincides
  with $g(\xone{y}{n})$, replace this equality by $g(\xone{x}{n})$.
  Otherwise, find the smallest number $i$ such that $x_i$ is different from
  $y_i$ and

  \begin{enumerate}
    \item add $y_i =_{\TA} x_i$ to $\Triang$;

    \item replace all occurrences of $y_i$ in $\Cha$ by $x_i$.
  \end{enumerate}
  We apply this transformation repeatedly until all equalities
  $g(\xone{x}{n}) =_{\TA} g(\xone{y}{n})$ disappear from the first row.

  So we can now assume that the first row contains no equalities and hence it
  has the form $q_n \KBe q_{n-1} \KBe \ldots \KBe q_1$, where all of the
  terms $q_i$ are quasi-flat.

  If all of the $q_i$ are variables, we can move $q_n \KBe q_{n-1} \KBe
  \ldots \KBe q_1$ to $\Simp$ and add $|q_1| > |t_1|$ to $\Arith$ obtaining
  an equivalent working constraint of smaller essential size. Hence, we can
  assume that at least one of the $q_i$ is a nonvariable term.

  Take any term $q_k$ in the first row such that $q_k$ is either a variable
  $x$ or a term $f^r(x)$. Note that other occurrences of $x$ in $\Cha$ can
  only be in the first row, and only in the terms of the form $f^k(x)$.

  Consider the formula $G$ defined as

    \begin{equation}
      \label{eqn:into:isolated:f}%
      \bigorl_{g \in \Sigma - \setof{f}}
      \bigorl_{m = 0 \ldots l}
       x =_{\TA} f^m(g(\bar{y})).
    \end{equation}
  where $\bar{y}$ is a sequence of pairwise different new variables. Since we
  proved that it is enough to restrict ourselves to solutions $\theta$ for
  which the height of $x\theta$ is at most $l$, the formulas $C$ and $C \andl
  G$ are equivalent up to $f$.

  Using the distributivity laws, $C \andl G$ can be turned into an equivalent
  disjunction of formulas $x =_{\TA} f^m(g(\bar{y})) \andl C$. For every such
  formula, replace $x$ by $f^m(g(\bar{y}))$ in the first row, and add $x
  =_{\TA} f^m(g(\bar{y}))$ to the triangle part. We do this transformation
  for all terms in the first row of the form $f^k(z)$, where $k \geq 0$ and
  $z$ is a variable. Now all the terms in the first row are of the form
  $f^m(g(\bar{y}))$, where $g$ is different from $f$ and $m \geq 0$.

  Let us show how to replace constraints of the first row with equivalent
  constraints consisting of constraints on variables and arithmetical
  constraints. Consider the pair $q_n,q_{n-1}$. Now $q_n =
  f^k(g(\xone{x}{u}))$ and $q_{n-1} = f^m(h(\xone{y}{v}))$ for some variables
  $\xone{x}{u},\xone{y}{v}$ and function symbols $g,h \in \Sigma -
  \setof{f}$. Then $q_n \KBe q_{n-1}$ is $f^k(g(\xone{x}{u})) \KBe
  f^m(h(\xone{y}{v}))$. If $k < m$ or ($k = m$ and $h \gg g$), then
  $f^k(g(\xone{x}{u})) \KBe f^m(h(\xone{y}{v}))$ is equivalent to $\false$.
  If $k > m$ or ($k = m$ and $g \gg h$), then $f^k(g(\xone{x}{u})) \KBe
  f^m(h(\xone{y}{v}))$ is equivalent to the arithmetical constraint
  $|g(\xone{x}{u})| =_{\nat} |h(\xone{y}{v})|$ which can be added to
  $\Arith$. If $k = m$ and $g = h$ (and hence $u=v$), then

    \[
      \begin{array}{r}
        f^k(g(\xone{x}{u})) \KBe f^m(h(\xone{y}{v}))
        \iffl
            |g(\xone{x}{u})| =_{\nat} |h(\xone{y}{v})|~\andl
            \\
            \displaystyle\bigorl_{i = 1 \ldots u}
              (x_1 =_{\TA} y_1 \andl \ldots \andl
                x_{i-1} =_{\TA} y_{i-1}~\andl x_i \KBo y_i).
      \end{array}
    \]
  We can now do the following. Add $|g(\xone{x}{u})| =_{\nat}
  |h(\xone{y}{v})|$ to $\Arith$ and replace $q_n \KBe q_{n-1}$ with the
  equivalent disjunction

    \[
      \displaystyle\bigorl_{i = 1 \ldots u}
        (x_1 =_{\TA} y_1 \andl \ldots \andl
          x_{i-1} =_{\TA} y_{i-1}~\andl x_i \KBo y_i).
    \]

%
  Then using the distributivity laws turn this formula into the equivalent
  disjunction of constraints of the form
    $
      C \andl x_1 =_{\TA} y_1 \andl \ldots \andl
      x_{i-1} =_{\TA} y_{i-1}~\andl x_i \KBo y_i
    $
  for all $i = 1 \ldots u$. For each of these constraints, we can move, as
  before, the equalities $x =_{\TA} y$ one by one to the triangle part
  $\Triang$, and make $\Cha \andl x_i \KBo y_i$ into a disjunction of chained
  constraints as in Lemma~\ref{lem:chained}.

  Let us analyze what we have achieved. After these transformations, in each
  member of the obtained disjunction the first row is removed from the
  chained part $\Cha$ of $C$. Since
  the row contained at
  least one function symbol, each member of the disjunction will contain at
  least one occurrence of a function symbol less than the original
  constraint. This is enough to prove termination of our algorithm, but not
  enough to present it as a nondeterministic polynomial-time algorithm. The
  problem is that, when $p_n$ is a variable $x$ or a term $f(x)$, one
  occurrence of $x$ in $p_n$ can be replaced by one or more
  constraints of the form  $x_i \KBo y_i$, where $x_i$
  and $y_i$ are new variables. To be able to show that the essential sizes of
  each of the resulting constraints is strictly less than the essential size
  of the original constraint, we have to modify our algorithm slightly.

  The modification will guarantee that the number of new variables introduced
  in the chained part of the constraint is not greater than the number of
  variables eliminated from the first row. We will achieve this by moving
  some constraints to the simple part $\Simp$.

  The new variables only appear when we replace a variable in the first row
  by a term $f^k(h(u_1,\ldots,u_m))$ or by $f^k(h(v_1,\ldots,v_m))$ obtaining a
  constraint  $f^k(h(u_1,\ldots,u_m))\linebreak[3] \KBe f^k(h(v_1,\ldots,v_m))$, which is
  then replaced by

    \begin{equation}
      \label{eqn:replacement}%
      u_1 =_{\TA} v_1 \andl \ldots \andl
      u_{i-1} =_{\TA} v_{i-1}~\andl u_i \KBo v_i.
    \end{equation}
  Let us call a variable $u_i$ (respectively, $v_i$) \emph{new} if
  $f^k(h(u_1,\ldots,u_m))$ occurred in the terms of the first row when we
  replaced a variable by a nonvariable term containing $h$ using formula
  \eqref{eqn:into:isolated:f}. In other words, new variables are those that
  did not occur in the terms of the first row before our transformation, but
  appeared in the terms of the first row during the transformation. All other
  variables are called \emph{old}. After the transformation we obtain a
  conjunction $E$ of constraints of the form $x_i =_{\TA} x_j$ or $x_i \KBo
  x_j$, where $x_i,x_j$ can be either new or old. Without loss of generality
  we can assume that this conjunction of constraints does not contain chains
  of the form $x_1 \# \ldots \# x_n \# x_1$ where $n \geq 2$ and at least one
  of the $\#$'s is $\KBo$. Indeed, if $E$ contains such a chain, then it is
  unsatisfiable.

  We will now show that the number of new variables can be restricted by
  moving constraints on these variables  into the triangle or simple part.
  Among the new variables, let us distinguish the following three kinds of
  variables. A new variable $x$ is called \emph{blue in $E$} if $E$ contains
  a chain $x =_{\TA} x_1 =_{\TA} \ldots =_{\TA} x_n$, where $x_n$ is an old
  variable. Evidently, a blue variable $x$ causes no harm since it can be
  replaced by an old variable $x_n$. Let us denote by
  \DI{$\oKB$}{$\protect\oKB$} the inverse relation to $\KBo$. A new variable
  $x$ is called \emph{red in $E$} if it is not blue in $E$ and $E$ contains a
  chain $x \# x_1 \# \ldots \# x_n$, where $x_n$ is an old variable, and all
  of the $\#$'s are either $=_{\TA}$, or $\KBo$, or $\oKB$. Red variables are
  troublesome, since there is no obvious way to get rid of them. However, we
  will show that the number of red variables is not greater than the number
  of replaced variables (such as the variable $x$ in
  \eqref{eqn:into:isolated:f}). Finally, all new variables that are neither
  blue nor red in $E$ are called \emph{green} in $E$.

  \textbf{Getting rid of the green variables.}
  We will now show that the green variables can be moved to the simple part
  of the constraint $\Simp$. To this end, note an obvious property: if $E$
  contains a constraint $x \# y$ and $x$ is green, then $y$ is green too. We
  can now do the following with the green variables. As in
  Lemma~\ref{lem:chained}, we can turn all the green variables into a
  disjunction of chained constraints of the form $v_1 \# \ldots \# v_n$,
  where $\#$ are $=_{\TA}$, $\KBw$, or $\KBe$, and use the distributivity
  laws to obtain chained constraints $v_1 \# \ldots \# v_n$. Let us call this
  constraint a \emph{green chain}. Then, if there is any equality $v_i
  =_{\TA} v_{i+1}$ in the green chain, we add this equality to $\Triang$ and
  replace this equality by $v_{i+1}$ in the chain. Further, if the chain has
  the form $v_1 \KBe \ldots \KBe v_k \KBw v_{k+1} \# \ldots \# v_{n}$, we add
  $v_1 \KBe \ldots \KBe v_k$ to $\Simp$ and $|v_k| > |v_{k+1}|$ to $\Arith$,
  and replace the green chain by $v_{k+1} \# \ldots \# v_{n}$. We do this
  transformation until the green chain becomes of the form $v_1 \KBe \ldots
  \KBe v_k$. After this, the green chain can be removed from $E$ and added to
  $\Simp$. Evidently, this transformation can be presented as a
  nondeterministic polynomial-time algorithm.

  \textbf{The red variables.} Let us show the following: in every term
  $f^k(h(\xone{u}{m}))$ in the first row at most one variable among
  $\xone{u}{m}$ is red. It is not hard to argue that it is sufficient to
  prove a stronger statement: if for some $i$ the variable $u_i$ is red or
  blue, then all variables $\xone{u}{i-1}$ are blue. So suppose that $u_i$ is
  either red or blue and $u_i \# y_n \# \ldots \# y_1$ is a shortest chain in
  $E$ such that $y_1$ is old. We prove that the variables $\xone{u}{i-1}$ are
  blue, by induction on $n$. When $n=1$ and $u_i$ is red, $E$ contains either
  $u_i \KBo y_1$ or $y_1 \KBo u_i$, where $y_1$ is old. Without loss of
  generality assume that $E$ contains $u_i \KBo y_1$. Then (cf.\
  \eqref{eqn:replacement}) this equation appeared in $E$ when we replaced
  $f^k(h(u_1,\ldots,u_m)) \KBe f^k(h(v_1,\ldots,v_m))$ by $u_1 =_{\TA} v_1
  \andl \ldots \andl u_{i-1} =_{\TA} v_{i-1}~\andl u_i \KBo v_i$ and $y_1 =
  v_i$. But then $E$ also contains the equations $u_1 =_{\TA} v_1 ,\ldots,
  u_{i-1} =_{\TA} v_{i-1}$, where the variables $\xone{v}{i-1}$ are old, and
  so the variables $\xone{u}{i-1}$ are blue. In the same way we can prove
  that if $u_i$ is blue then $\xone{u}{i-1}$ are blue. The proof for $n >1$
  is similar, but we use the fact that $\xone{v}{i-1}$ are blue rather than
  old.

  To complete the transformation, we add all constraints on the red and the
  old variables to $\Cha$ and make $\Cha$ into a disjunction of chained
  constraint as in Lemma~\ref{lem:chained}.

  \textbf{Getting rid of the blue variables.} If $E$ contains a blue variable
  $x$, then it also contains a chain of constraints $x =_{\TA} x_1 =_{\TA}
  \ldots =_{\TA} x_n$, where $x_n$ is an old variable. We replace $x$ by
  $x_n$ in $C$ and add $x =_{\TA} x_n$ to the triangle part $\Triang$.

  When we completed the transformation on the first row, the row disappears
  from the chained part $\Cha$ of $C$. If the first row contained no function
  symbols, the size of $\Cha$ will become smaller, since several variables
  will be removed from it. If $\Cha$ contained at least one function symbol,
  that after the transformation the number of occurrences of function symbols
  in $\Cha$ will decrease. Some red variables will be introduced, but we
  proved that their number is not greater than the number of variables
  eliminated from the first row. Therefore, the size of $\Cha$ strictly
  decreases after the transformation due to elimination of at least one
  function symbol.

Again, it is not hard to argue that the transformation can be presented as a
nondeterministic polynomial-time algorithm computing all members of the
resulting disjunction of constraints.

\end{proof}

Lemmas~\ref{lem:chained} and \ref{lem:into:isolated} imply the following:

\begin{lemma}
\label{lem:into:isolated:strong}%
  Let $C$ be a constraint. Then there exists a disjunction $C_1 \orl \ldots
  \orl C_n$ of constraints in isolated form equivalent to $C$ up to $f$.
  Moreover, members of such a disjunction can be found by a nondeterministic
  polynomial-time algorithm.
\end{lemma}

Our next aim is to present a nondeterministic polynomial-time algorithm
solving constraints in isolated form.


       \section{From constraints in isolated form to systems of linear
                           Diophantine inequalities}
                       \label{sec:isolated->diophantine}

Let $C$ be a constraint in isolated form

  \[
    \Simp \andl \Arith \andl \Triang.
  \]
Our decision algorithm will be based on a transformation of the simple
constraint $\Simp$ into an equivalent disjunction $D$ of arithmetical
constraints. Then we can check the satisfiability of the resulting formula $D
\andl \Arith$ by using an algorithm for solving systems of linear Diophantine
inequalities on the weights of variables.

To transform $\Simp$ into an arithmetical formula, observe the following. The
constraint $\Simp$ is a conjunction of the constraints of the form

  \[
    x_1 \KBe \ldots \KBe x_N
  \]
having no common variables. To solve such a constraint we have to ensure that
there exist at least $N$ different terms of the same weight as $x_1$ (since
the Knuth-Bendix order is total).

In this section we will show that for each $N$ the statement ``there exists
at least $N$ different terms of a weight $w$'' can be expressed in the
Presburger Arithmetic as an existential formula of one variable $w$.

We say that a relation $R(\bar{x})$ on natural numbers is
\DI{$\exists$-definable}{$\exists$-definable}, if there exists an existential
formula of Presburger Arithmetic $C(\bar{x},\bar{y})$ such that
$R(\bar{x})$ is equivalent to $\exists\bar{y} C(\bar{x},\bar{y})$. We call a
function $r(\bar{x})$ $\exists$-definable if so is the relation $r(\bar{x}) =
y$. Note that composition of $\exists$-definable function is
$\exists$-definable.

Let us fix an enumeration $\xone{g}{S}$ of the signature $\Sigma$. We assume
that the first $B$ symbols $\xone{g}{B}$ have an arity $\geq 2$, and the first
$F$ symbols $\xone{g}{F}$ are nonconstants. The arity of each $g_i$ is
denoted by $\arity_i$. \emph{In this section we assume that $B$, $F$, $S$,
and the weight function $w$ are fixed.}

We call the \DI{contents}{contents} of a ground term $t$ the tuple of natural
numbers $(\xone{n}{S})$ such that $n_i$ is the number of occurrences of $g_i$
in $t$ for all $i$. For example, if the sequence of elements of $\Sigma$ is
$g,h,a,b$, and $t = h(g(h(h(a)),g(b,b)))$, the contents of $t$ is
$(2,3,1,2)$.

\begin{Lemma}
\label{lem:exists}%
  The following relation $\Exists(x,\xone{n}{S})$ is
  $\exists$-definable: there exists at least one ground term of $\Sigma$ of
  the weight $x$ and contents $(\xone{n}{S})$.
\end{Lemma}
\begin{proof}
  We will define $\Exists(x,\xone{n}{S})$ by a conjunction of two
  linear Diophantine inequalities.

  The first equation is

    \begin{equation}
    \label{eqn:signature1}%
      x = \sum_{1\leq i \leq S} w(g_i) \cdot n_i.
    \end{equation}
  It is not hard to argue that this equation says: every term with the
  contents $(\xone{n}{S})$ has weight $x$.

  The second formula says that the number of constant and nonconstant
  function symbols in $(\xone{n}{S})$ is appropriately balanced for
  constructing a term:

    \begin{equation}
    \label{eqn:signature2}%
      1 + \sum_{1 \leq i \leq S}(\arity_i-1)\cdot n_i=0.
    \end{equation}
\end{proof}

Let us prove some lower bounds on the number of terms of a fixed weight.

We leave the following two lemmas to the reader. The first one implies that,
if there exists any ground term $t$ of a weight $x$ with at least $N$
occurrences of nonconstant symbols, including at least one occurrence of a
function symbol of an arity $\geq 2$, then there exists at least $N$ different
ground terms of the weight $x$.

\begin{lemma}
\label{lem:N:binary}%
  Let $x,\xone{n}{S}$ be natural numbers such that $\Exists(x,\xone{n}{S})$
  holds, $n_1 + \ldots + n_B \geq 1$ and $n_1 + \ldots + n_F \geq N$. Then
  there exists at least $N$ different ground terms with the contents
  $(\xone{n}{S})$.
\end{lemma}

The second lemma implies that, if there exists any ground term $t$ of a
weight $x$ with at least $N$ occurrences of nonconstant function symbols,
including at least two different unary function symbols, then there exists at
least $N$ different ground terms of the weight $x$.

\begin{lemma}
\label{lem:N:unary}%
  Let $x,\xone{n}{s}$ be natural numbers such that $\Exists(x,\xone{n}{S})$
  holds, $n_1 +\ldots + n_F \geq N$ and at least two numbers among
  $n_{B+1},\ldots,n_F$ are positive. Then there exists at least $N$ different
  ground terms with the contents $(\xone{n}{S})$.
\end{lemma}
Let us note that if our signature consists only of a unary function symbol of
a positive weight and constants, then the number of different terms in any
weight is less or equal to the number of constants in the signature.

The remaining types of signatures are covered by the following lemma.

\begin{Lemma}
\label{lemm:N:linear}
  Let  $\Sigma$ contain a function symbol of an arity greater than or equal to
  $2$, or contain at least two different unary function symbols. Then there
  exist two natural numbers $N_1$ and $N_2$ such that for all natural numbers
  $N$ and $x$ such that $x > N \cdot N_1+N_2$, the number of terms of the
  weight $x$ is either $0$ or greater than $N$.
\end{Lemma}
\begin{Proof}
  If $\Sigma$ contains a unary function symbol of the weight $0$ then the
  number of different terms of any weight is either $0$ or $\omega$ and the
  lemma trivially holds.

  Therefore we can assume that our signature contains no unary function
  symbol of the weight $0$. Define

    \[
      \begin{array}{rcl}
        W &=& \max \{w(g_i)|1\leq i\leq S\}; \\
        A &=& \max \{\arity_i|1\leq i\leq S\}; \\
        N_1 &=& W \cdot A ; \\
        N_2 &=& W^2 \cdot (A + 1) + W.
      \end{array}
    \]
  Take any $N$ and $x$ such that $x > N\cdot N_1 + N_2$.

  Let us prove that if there exists a term of the weight $x$ then the number
  of occurrences of nonconstant function symbols in this term is greater than
  $N$. Assume the opposite, i.e.\ there exists a term $t$ of the weight $x$
  such that the number of occurrences of nonconstant function symbols in $t$
  is $M \leq N$. Let $(\xone{n}{S})$ be the contents of $t$ and $L$ denote
  the number of occurrences of constants in $t$. Note that
  \eqref{eqn:signature2} implies $L = 1 + \sum_{1\leq i\leq F}(\arity_{i} -
  1) \cdot n_i$. Then using \eqref{eqn:signature1} we obtain

    \[
      \begin{array}{l}
        N\cdot N_1 + N_2 <
        |t| =
        \sum_{1\leq i \leq S} w(g_i) \cdot n_i  \leq
        W \cdot \sum_{1\leq i \leq S} \cdot n_i  =
        \\[.9ex]
        W\cdot(M+L) =
        W\cdot(M + 1 + \sum_{1\leq i\leq F}(\arity_{i} - 1)\cdot n_i) \leq
        \\[.9ex]
        W\cdot(M + 1 + (A-1)\sum_{1\leq i\leq F} n_i) =
        W\cdot(M+1 + (A-1) \cdot M ) =
        \\[.9ex]
        W\cdot(M\cdot A +1) \leq
        W\cdot(N \cdot A +1) <
        N\cdot N_1+N_2 .
      \end{array}
    \]
 So we obtain a contradiction.


  Consider the following possible cases.

  \begin{enumerate}
    \item \emph{There exists a term of the weight $x$ with an occurrence of a
    function symbol of an arity greater than or equal to $2$.} In this case
    by Lemma \ref{lem:N:binary} the number of different terms of the weight
    $x$ is greater than $N$.

    \item \emph{There exists a term of the weight $x$ with occurrences of at
    least two different unary function symbols}. In this case by Lemma
    \ref{lem:N:unary} the number of different terms of the weight $x$ is
    greater than $N$.

    \item \emph{All terms of the weight $x$ have the form $g^k(c)$ for some
    unary function symbol $g$ and a constant $c$.} We show that this case is
    impossible. In particular, we show that for any nonconstant function
    symbol $h$ there exists a term of the weight $x$ in which $g$ and $h$
    occur, therefore we obtain a contradiction with the assumption.

    We have $x=w(g)\cdot k + w(c)$. Denote by $H$ the arity of $h$. Let us
    define integers $M_1,M_2,M_3$ as follows

      \[
        \begin{array}{rcl}
          M_1 &=& w(g), \\
          M_2 &=& k - w(h) - w(c) \cdot(H - 1), \\
          M_3 &=& w(g) (H - 1) + 1.
        \end{array}
      \]
    Let us prove that $M_1,M_2,M_3 > 0$ and there exists a term of the weight
    $x$ with $M_1$ occurrences of $h$, $M_2$ occurrences of $g$ and $M_3$
    occurrences of $c$ and hence obtain a contradiction.

    Since $g$ is unary, $w(g) > 0$, and so $M_1 > 0$. Since $H \geq 1$, we
    have $M_3 > 0$. Let us show that $M_2 > 0$, i.e.\
    $k>w(h)+w(c)\cdot(H-1)$. We have

      \[
        \begin{array}{l}
          k = (x-w(c)) / w(g) >
          (N\cdot N_1 + N_2 - w(c)) / w(g) \geq \\
          (N_2 - w(c)) / w(g) =
          (W^2\cdot(A+1)+W-w(c)) / w(g) \geq \\
          (W^2\cdot(A+1)) / w(g) \geq
          W \cdot(A+1) =
          W + W \cdot A \geq \\
          w(h) + w(c) \cdot A >
          w(h) + w(c)\cdot(H-1) .
        \end{array}
      \]
    It remains to show that there exists a term of the weight $x$ with $M_1$
    occurrences of $h$, $M_2$ occurrences of $g$ and $M_3$ occurrences of
    $c$. To this end we have to prove (cf.\ \eqref{eqn:signature1} and
    \eqref{eqn:signature2})

      \[
        \begin{array}{l}
          x = w(h) \cdot M_1 + w(g) \cdot M_2 + w(c) \cdot M_3,
          \\[1ex]
          1 + (H - 1) \cdot M_1 + (1 - 1) \cdot M_2 + (0 - 1) M_3 = 0.
        \end{array}
      \]
    This equalities can be verified directly by replacing $M_1,M_2,M_3$ by
    their definitions and $x$ by $w(g)\cdot k + w(c)$. $\QED$
  \end{enumerate}
\end{Proof}

Define the binary function \DI{$\TT$}{tnt@$\protect\TT$} (truncated number of
terms) as follows: $\TT(N,M)$ is the minimum of $N$ and the number of terms
of the weight $M$ and let us show that $\TT$ can be computed in time
polynomial of $N+M$. To give a polynomial-time algorithm for this function we
need an auxiliary definition and a lemma.

\begin{definition}
  Let $(n_1,\ldots,n_s)$ and $(m_1,\ldots,m_s)$ be two tuples of natural
  numbers. We say that $(n_1,\ldots,n_s)$ \DI{extends}{extends}
  $(m_1,\ldots,m_s)$ if $n_i\geq m_i$ for $1\leq i \leq s$.
\end{definition}

The \DI{depth}{depth} of a term is defined by induction as usual: the depth
of every constant is $1$ and the depth of every nonconstant term
$g(\xone{t}{n})$ is equal to the maximum of the depth of the $t_i$'s plus
$1$.

\begin{Lemma}
 \label{lem:extended:content}
   Let $t_1,\ldots,t_n$ be a collection of different terms of the same depth
   and $\Con$ be the contents of a term such that $\Con$ extends the contents
   of all terms $t_i$, $1\leq i \leq n$. Then there exists at least $n$
   different terms with the contents $\Con$.
\end{Lemma}
\begin{proof}
  Let us define the notion of \emph{leftmost subterm} of a term $t$ as
  follows: every constant $c$ has only one leftmost subterm, namely $c$
  itself, and leftmost subterms of a nonconstant term $g(\xone{r}{n})$ are
  this term itself and all leftmost subterms of $r_1$. Evidently, for each
  positive integer $d$ and term $t$, $t$ has at most one leftmost subterm of
  the depth $d$.

  It is not hard to argue that from the condition of the lemma it follows
  that for every  term $t_i$ there exists a term $s_i$ with the contents
  $\Con$ such that $t_i$ is a leftmost subterm of $s_i$. But then the terms
  $\xone{s}{n}$ are pairwise different, since they have different leftmost
  subterms of the depth $d$.
\end{proof}

\begin{Lemma}
  \label{lem:NumbNM}%
  Let the signature $\Sigma$ contain no unary function symbol of the weight
  $0$ and contain either a function symbol of an arity greater than or equal
  to $2$ or contain at least two different unary function symbols. Then the
  function $\TT(N,M)$ is computable in time polynomial of $M + N$.
\end{Lemma}
\begin{proof}
  It is not hard to argue that for every contents $(\xone{n}{S})$ such that
  some of the $n_i$'s is greater than $M$, any term with these contents has
  the weight greater than $M$. The number of different contents in which each
  of the $n_i$'s is less or equal than $M$ is $M^S$, i.e.\ it is polynomial
  in $M$, moreover, all these contents can be obtained by an algorithm
  working in time polynomial in $M$.

  Therefore it is sufficient to describe a polynomial-time algorithm which
  for all contents $(n_1,\ldots,n_S)$, where $1\leq n_i\leq M$, returns the
  minimum of $N$ and the number of terms with these contents.

  Let us fix contents $\mathit{Con}=(n_1,\ldots\,n_S)$ where $1\leq n_i\leq
  M$. Using equations \eqref{eqn:signature1} and \eqref{eqn:signature2}, one
  can check in polynomial time whether there exists a term with the contents
  $\Con$, so we assume that there exists at least one such term.

  Our algorithm constructs, step by step, sets $T_0,T_1,\ldots$, of different
  terms with contents which can be extended to the contents $\mathit{Con}$.
  Each set $T_i$ will consist only of terms of the depth $i$.

  \begin{enumerate}
    \item \emph{Step 0.} Define $T_0 = \emptyset$.

    \item \emph{Step $i+1$.} Define

      \[
        T_{i+1} = \{
                    g(\xone{t}{m}) \mid
                      \begin{array}[t]{l}
                        g \in \Sigma,~
                          \xone{t}{m} \in T_1 \cup \ldots \cup T_i,~\\
                          \mathit{Con} \text{ extends the content of }
                            g(\xone{t}{m}), \text{ and}\\
                          \text{the depth of $g(\xone{t}{m})$ is } i + 1
                        \}.
                      \end{array}
      \]
      If $T_{i+1}$ has $N$ or more terms, then by
      Lemma~\ref{lem:extended:content} there exists at least $N$ different
      terms of the content $\Con$, so we terminate and return $N$. If
      $T_{i+1}$ is empty, we return as the result the minimum of $N$ and the
      number of terms with the content $\Con$ in $T_1 \cup \ldots \cup
      T_{i+1}$.
  \end{enumerate}
  Let us prove some obvious properties of this algorithm.

  \begin{enumerate}
    \item \emph{If some $T_i$ contains $N$ or more terms, then there exists
    at least $N$ terms with the content $\Con$.} As we noted, this follows
    from Lemma~\ref{lem:extended:content}.

    \item \emph{At the end of step $i+1$ the set $T_1 \cup \ldots T_{i+1}$
    contains all the terms with the contents $\Con$ of the depth $\leq i+1$.}
    This property obviously holds by our construction.
  \end{enumerate}
  This property ensure that the algorithm is correct. To prove that it works
  in time polynomial in $M+N$ it is enough to note that each step can be made
  in time polynomial in $N$ and the total number of steps is at most $M+1$.
\end{proof}

Now we are ready to prove
the main lemma of this section.

\begin{Lemma}
  \label{lem:atleast}%
   There exists a polynomial time of $N$ algorithm,  which constructs an
   existential formula $\atleast_N(x)$ valid on a natural number $x$ if and
   only if there exists at least $N$ different terms of the weight $x$.
\end{Lemma}
\begin{proof}
  If the signature $\Sigma$ contains a unary function symbol of the weight
  $0$ then the number of different terms in any weight is either $0$ or
  $\omega$. Therefore we can define $\atleast_N(x)$ as $\exists n_1 \ldots
  \exists n_S \Exists(x,\xone{n}{S})$.

  Let us consider the case when the signature $\Sigma$ consists of a unary
  function symbol $g$ of a positive weight. For every constant $c$ in
  $\Sigma$ consider the formula $G_c(x) = \exists k(w(g)k + w(c) = x)$. It is
  not hard to argue that $G_c(x)$ holds if and only if there exists a term of
  the form $g^k(c)$. Let $P$ be the set of all sets of cardinality $N$
  consisting of constants of $\Sigma$ (the cardinality of $P$ is obviously
  polynomial in $N$). It is easy to see that

    \[
      \atleast_N(x)\iffl \bigorl_{Q\in P}\bigandl_{Q \in S} G_c(x).
    \]
  It  remains to consider the case when our signature contains a function
  symbol of an arity greater than or equal to $2$, or contain at least two
  different unary function symbols. By Lemma \ref{lemm:N:linear}, there exist
  constants $N_1$ and $N_2$ such that for any natural number $x$ such that
  $x > N \cdot N_1+N_2$ the number of terms of the weight $x$ is either $0$
  or greater than $N$. Let us denote $N \cdot N_1+N_2$ as $M$ and the set
  $\{M'|M'\leq M \andl \TT(N,M')\geq N\}$ as $W$. By Lemmas
  \ref{lemm:N:linear}, \ref{lem:NumbNM} we have

    \[
       \atleast_N(x) \iffl
          (\exists \xone{n}{S}
              \Exists(x,\xone{n}{S})~\andl x > M)\bigorl
               (\bigorl_{M'\in W} x=M').
   \]
\end{proof}

                           \section{Main results}
                           \label{sec:main:result}


In this section we complete the proofs of the main results of this \paper.

\begin{Theorem}
\label{thm:KBOisNP}%
  For every Knuth-Bendix order, the problem of solving ordering constraints
  is contained in NP.
\end{Theorem}
\begin{proof}
  Take a constraint. By Lemma~\ref{lem:into:isolated} it can be effectively
  transformed into an equivalent disjunction of isolated forms, so it remains
  to show how to check satisfiability of constraints in isolated form.

  Suppose that $C$ is a constraint in isolated form. Recall that $C$ is of
  the form

    \begin{equation}
    \label{eqn:main:1}
      \Arith \andl \Triang \andl \Simp.
    \end{equation}

  Let $\Simp$ contain a chain $x_1 \KBe \ldots \KBe x_N$ such that
  $\xone{x}{N}$ does not occur in the rest of $\Simp$. Denote by $\Simp'$ the
  constraint obtained from $\Simp$ by removing this chain. It is not hard to
  argue that $C$ is equivalent to the constraint

    \[
        \Arith \andl \Triang \andl \Simp' \andl
        \displaystyle\bigandl_{i = 2 \ldots N} (|x_i| =_{\nat} |x_1|) \andl
        \atleast_{N}(|x_1|).
    \]
  In this way we can replace $\Simp$ by an arithmetical constraint, so we
  assume that $\Simp$ is empty. Let $\Triang$ have the form

    \[
    y_1 =_{\TA} t_1 \andl \ldots \andl y_n =_{\TA} t_n.
    \]
  Let $Z$ be the set of all variables occurring in $\Arith \andl \Triang$. It
  is not hard to argue that $\Arith \andl \Triang$ is satisfiable if and only
  if the following constraint is satisfiable:

    \[
      \begin{array}{l}
        \Arith \andl |y_1| =_{\nat} |t_1| \andl \ldots \andl |y_n| =_{\nat} |t_n| \andl
        \bigandl_{z \in Z} \atleast_{1}(|z|).
      \end{array}
    \]
  So we reduced the decidability of the existential theory of term algebras
  with a Knuth-Bendix order to the problem of solvability of systems of
  linear Diophantine inequalities. Our proof can be represented as a
  nondeterministic polynomial-time algorithm.

\end{proof}

This theorem implies the main result of this \paper. Let us call a signature
$\Sigma$ \emph{trivial} if it consists of one constant symbol. Evidently, the
first-order theory of the term algebra of a trivial signature is polynomial.

\begin{Theorem}
\label{thm:main}%
  The existential first-order theory of any term algebra of a non-trivial
  signature with the Knuth-Bendix order is NP-complete.
\end{Theorem}
\begin{proof}
  The containment in NP follows from Theorem~\ref{thm:KBOisNP}. It is easy to
  prove NP-hardness by reducing propositional satisfiability to the
  existential theory of the algebra (even without the order).
\end{proof}

Let us show that for some Knuth-Bendix orders even constraint solving can be
NP-hard.

\begin{example}
\label{exa:LDI}%
  Consider the signature $\Sigma = \setof{s,g,h,c}$, where $h$ is binary,
  $s,g$ are unary, and $c$ is a constant. Define the weight of all symbols as
  $1$, and use any order $\gg$ on $\Sigma$ such that $g \gg s$. Our aim is
  to represent any linear Diophantine equation by Knuth-Bendix constraints.
  To this end, we will consider any ground term $t$ as representing the
  natural number $|t|-1$.

  Define the formula

    \[
      \begin{array}{l}
        \mathit{equal\_weight}(x,y) \iffl \\ ~~~~~~~~~
          g(x) \KBo s(y) \andl g(y) \KBo s(x).
      \end{array}
    \]
  It is not hard to argue that, for any ground terms $r,t$
  $\mathit{equal\_weight}(r,t)$ holds if and only if $|r|=|t|$.

  It is enough to consider systems of linear Diophantine equations of the
  form

    \begin{equation}
      \label{eqn:Dio}
      x_1 + \ldots + x_n + k = x_0,
    \end{equation}
  where $x_0,\ldots,x_n$ are pairwise different variables, and $k \in \nat$.
  Consider the constraint

    \begin{equation}
    \setlength{\arraycolsep}{0pt}
      \label{eqn:Con}
        \mathit{equal\_weight}(
          \begin{array}[t]{l}
            s^{k+2}(
                  \begin{array}[t]{l}
                      h(y_1,h(y_2,\ldots, \\
                      ~~~~~h(y_{n-1},y_n)))),
                  \end{array} \\
            s^{2n}(y_0)).
          \end{array}
    \end{equation}
  It is not hard to argue that

  \begin{eqitemize}
  \item Formula \eqref{eqn:Con} holds if and only if
  \label{kkk}

    \[
       |y_1| - 1 + \ldots + |y_n| - 1 + k = |y_0| - 1.
    \]
  \end{eqitemize}
  Using \eqref{kkk}, we can transform any system $D(\xone{x}{n})$ of linear
  Diophantine equations of the form \eqref{eqn:Dio} into a constraint
  $C(\xone{y}{n})$ such that for every tuple of ground terms $\xone{t}{n}$,
  $C(\xone{t}{n})$ holds if and only if so does $D(|t_1|-1,\ldots,|t_n|-1)$.
 
 Similar, using a formula 
    \[
      \begin{array}{l}
        \mathit{greater\_weight}(x,y) \iffl \\ ~~~~~~~~~
          s(x) \KBo g(y) 
      \end{array}
    \]
 one can represent systems of linear inequalities using Knuth--Bendix constraints.
\end{example}

Since it is well-known that solving linear Diophantine equations is
NP-hard, we have the following theorem.

\begin{theorem}
\label{thm:KBOisNPcomp}%
  For some Knuth-Bendix orders, the problem of solving ordering constraints
  is NP-complete.
\end{theorem}

This result does not hold for all non-trivial signatures, as the following
theorem shows.

\begin{Lemma} 
    There exists a polynomial time  algorithm which solves ordering constraints 
   for  any given term algebra over a signature consisting of constants 
   and any total ordering $\succ$ on that term algebra.
\end{Lemma}
\begin{proof}
  Let $\Sigma=\{c_1,\ldots,c_n\}$, w.l.o.g. we can assume that 
  $c_n\succ c_{n-1}\succ\ldots\succ c_1$. 
  Let $C$ be an ordering constraint. First we get rid of equalities
  as follows. If $t=_{\TA}s$ occurs in $C$  and  $t$ syntactically
  equal to $s$ then we remove  $t=_{\TA}s$ from $C$, if 
  $t$  is  a variable then we replace all occurrences of $t$ in $C$ by $s$ 
  and remove $t=_{\TA}s$ from $C$, 
  otherwise $t$ and $s$ are different constants and $C$ is unsatisfiable.
  Now $C$ consists of conjunctions of atomic formulas 
  of the form $t\succ s$. 
  We define a relation $\succ_{C}'$ on terms as follows: $t\succ_{C}' s$ if and only if 
  $t \succ s$ occurs into $C$. Let $\succ_{C}$  denote a transitive
  closure of $\succ_{C}'$. It is easy to see, that  using a polynomial
  time algorithm for
  transitive closure, we can compute the relation 
  $t\succ_{C} s$ in polynomial time. Note that if $\succ_{C}$ is
  not a strict order then the constraint $C$ is unsatisfiable.
  So we assume that $\succ_{C}$ is a strict partial order.  
 
 Now we replace all variables in $C$ by constants as follows.
 Take a variable $x$ such that there is no 
  variable less than $x$ w.r.t. $\succ_{C}$.
 There are two possible cases: 
\begin{enumerate}  
 \item \label{lem:const:min}
  $x$ is a minimal term w.r.t. $\succ_{C}$, then we replace 
       all occurrences of $x$ in $C$ by $c_1$. 
\item \label{lem:const:const}
 there exist some constants less than  $x$ w.r.t. $\succ_{C}$, then 
  let $c_{\mathit{max}}$ be the greatest w.r.t. $\succ$
  constant among such constants. 
  If $c_{\mathit{max}}$ is the
  maximal constant in $\TA(\Sigma)$ then the constraint $C$ is unsatisfiable, otherwise
  we replace all occurrences of $x$ by $c_{\mathit{max}+1}$.
\end{enumerate}
  Repeating this process we replace all variables in $C$ 
 in polynomial time.
 To complete the proof of the lemma, it  remains to  show that transformations
  \ref{lem:const:min},\ref{lem:const:const} above, preserve
  satisfiability of constraints without equality. 
 To this end, we consider a constraint $C$ without equality 
  and a solution $\theta$ to $C$. 
  If the  transformation \ref{lem:const:min} is applicable to $C$ 
  then it is easy to see that 
 \[\theta'(x) = \left\{
 \begin{array}{l}
  \text{$c_1$, if $x$ is a minimal w.r.t. $\succ_{C}$,}\\
  \text{$\theta(x)$ otherwise.}
  \end{array}    
  \right.
 \] 
 is a solution to the constraint obtained after applying the
transformation \ref{lem:const:min} to $C$.

Similar one can show that the transformation \ref{lem:const:const}
preserves satisfiability of constraints without equality.

\end{proof}


\begin{corollary} \label{ex:realKBO}
 There exists a polynomial time algorithm which checks solvability of
 ordering  constraints 
  for any given Knuth--Bendix order on any  term algebra over a
signature consisting of constants.
\end{corollary}
 
As we mentioned in Section \ref{sec:preliminaries}, 
if we consider  real--valued Knuth-Bendix orders then 
even comparison of ground terms might be undecidable. 
Let us show it on the following example. 

\begin{example} Consider a non-computable real number 
 $r$ such that $0<r < 1$, i.e. 
  there is no algorithm which given 
 a positive integer $n$ computes $r$ with the precision $1/n$, 
 in other words finds two natural numbers $p,q$ such that 
 $|r-p/q| < 1/n$. 
  
 Now we consider a signature consisting of two unary symbols 
 $g,h$ and a constant $c$ and consider any Knuth--Bendix order $\KBo$
on the corresponding term algebra, such that $w(g)=1$ and $w(h)=r$.
%
Let us show that comparison of terms in this Knuth--Bendix
 order is undecidable.
 Consider a positive integer $n$. Then, it is easy to see 
 that  there exists a positive integer
 $m$ such that $g^m(c)\KBo h^n(c)\KBo g^{m-1}(c)$. 
 Since $|g^m(c)| \not = | h^n(c)|\not = |g^{m-1}(c)|$, we have  $|g^m(c)|>|h^n(c)|> |g^{m-1}(c)|$.
 From the definition of the weight function we have that 
 $m>rn >  m-1$ and therefore $m/n>r >  \frac{m-1}{n}$. 
 Let us take $p=m-1$ and $q=n$, then we have $|r-p/q|< 1/n$.  
 Therefore using comparison of terms  we can compute $r$ with the precision
 $1/n$. This implies that  comparison of terms for this Knuth--Bendix order is undecidable.   
\end{example}


                  \section{Related work and open problems}
                             \label{sec:related}

In this section we overview previous work on Knuth-Bendix orders, recursive
path orders, and extensions of term algebras with various relations.

The Knuth-Bendix order was introduced in
\cite{KnuthBendix:Pergamon:WordProblems:1970}. Later, \Cite{der82} introduced
recursive path orders (RPOs) and \Cite{KaminLevy:LPO:1980} lexicographic path
orders (LPOs). A number of results on recursive path orders and solving LPO
and RPO ordering constraints are known.

However, except for the very general result of \cite{nie93} the techniques
used for RPO constraints are not directly applicable to Knuth-Bendix orders.
We used systems of linear Diophantine inequalities in our decidability
proofs. This is not coincidental: Example~\ref{exa:LDI} shows that systems
of linear Diophantine inequalities are definable in the Knuth-Bendix order.

\Cite{ComonTreinen:CAAP:OrderConstr:1994} proved that LPO constraint solving
is NP-hard already for constraints consisting of a single inequality. In
\cite{KorovinVoronkov:RTA:KBorientabilityIsNP:2001} we prove that the problem
of solving Knuth-Bendix ordering constraints consisting of a single
inequality can be solved in polynomial time.

In  \cite{KorovinVoronkov:RTA:KBorientabilityIsNP:2001} we 
present a polynomial time algorithm for the orientability problem:
given a system of rewrite rules
$R$, does there exist a  Knuth--Bendix order which orients every ground
instance of every rewrite rule in $R$.
A similar problem of orientability by the non-ground version of the
real--valued Knuth--Bendix order was studied by Dick, Kalmus, and Martin
\cite{mar87,Dick+:ActaI:KBO:1990} and an algorithm for orientability was given.
Algorithms for, and complexity of, orientability problem for various versions of the
recursive path orders were considered in
\cite{Lescanne:CADE:orientability:1984,%
DetlefsForgaard:RTA:ProcedureLPO:1985,%
KrishnamoorthyNarendran:TCS:RPO:1985}. 
In particular, in \cite{KrishnamoorthyNarendran:TCS:RPO:1985}
it is shown that the orientability problem by the  non-ground version
of the recursive path order is NP-complete.



\Cite{Comon:JFCS:OrderingConstraints:1990} proved the decidability and
\Cite{nie93} NP-completeness of LPO constraint solving.
\Cite{JouannaudOkada:ICALP:SubtermConstraints:1991} proved the decidability
and \Cite{Narendran+:CSL:RPOinNP:1999} NP-completeness of RPO constraint
solving. Recently, \Cite{NieuwenhuisRivero:RTA:SolvedFormsLPO:1999} proposed
a new efficient method for solving  RPO constraints.

\Cite{Lepper:TCS:KBOrderType:2001} studies derivation length and order types
of Knuth-Bendix orders, both for integer-valued and real-valued weight
functions.

Term algebras are rather well-studied structures.
\Cite{Malcev:Doklady:TermAlgebras:1961} was the first to prove the
decidability of the first--order theory of term algebras. Other methods of
proving decidability were developed by
\Cite{ComonLescanne:JSC:Disunification:1989},
\Cite{Kunen:JLP:NegationLP:1987},
\Cite{Belegradek:TrudyIM:LocallyFreeAlg:1988},
\Cite{Maher:LICS:AxiomTermAlgebra:1988}.

If we introduce a  binary predicate into a term algebra, then one can obtain
a richer theory. Term algebras with the subterm predicate have an undecidable
first--order theory and a decidable existential theory
\cite{Venkataraman:JACM:TAwithSubterm:1987}. Term algebras with lexicographic
path orders have an undecidable first--order theory
\cite{ComonTreinen:TCS:FO-LPO-undec:1997}. However, if we consider 
term algebras over signatures consisting of unary symbols and
constants then the first--order theory of lexicographic
path orders over such term algebras is decidable \cite{Narendran:Rusinowitch:FO:UnaryRPO}. 
In \cite{KV:FO:KBO}  we show that the first--order theory of any  Knuth--Bendix 
order over any term algebra over a signature consisting of unary 
function symbols and constants is decidable.

To conclude, we  mention two open problems related to the
Knuth--Bendix order.
One problem is whether whole first--order theory of the Knuth--Bendix
orders is decidable. 
Another problem is to describe   the complexity of the 
constraint solving problem for Knuth--Bendix orders 
in the case of signatures consisting of unary function symbols and constants.




\begin{small}

\begin{theindex}

\addcontentsline{toc}{section}{Index}
{\bfseries\hfil\normalsize Symbols\hfil}\nopagebreak

  \item $=_{\TA }$~~\dotfill 4
  \item $=_{\nat }$~~\dotfill 4
  \item $\KBe $~~\dotfill 5
  \item $\KBo $ --- Knuth Bendix order~~\dotfill 3
  \item $\KBw $~~\dotfill 5
  \item $\exists $-definable~~\dotfill 16
  \item $\false $~~\dotfill 6
  \item $\mid t \mid $ --- weight of $t$~~\dotfill 2
  \item $\oKB $~~\dotfill 14

  \indexspace
{\bfseries\hfil\normalsize A\hfil}\nopagebreak

  \item arithmetical sort~~\dotfill 4

  \indexspace
{\bfseries\hfil\normalsize C\hfil}\nopagebreak

  \item chained constraint~~\dotfill 5
  \item compatible precedence relation~~\dotfill 2
  \item constant~~\dotfill 2
  \item constraint~~\dotfill 4
  \item contents~~\dotfill 16

  \indexspace
{\bfseries\hfil\normalsize D\hfil}\nopagebreak

  \item dependent~~\dotfill 7
  \item depth~~\dotfill 18

  \indexspace
{\bfseries\hfil\normalsize E\hfil}\nopagebreak

  \item equivalence~~\dotfill 4
  \item equivalent up to $f$~~\dotfill 9
  \item essential size~~\dotfill 8
  \item extends~~\dotfill 18

  \indexspace
{\bfseries\hfil\normalsize F\hfil}\nopagebreak

  \item $f$~~\dotfill 2
  \item $f$-difference~~\dotfill 9
  \item $f$-height~~\dotfill 9
  \item $f$-term~~\dotfill 9
  \item $f$-variant~~\dotfill 9
  \item flat term~~\dotfill 5

  \indexspace
{\bfseries\hfil\normalsize G\hfil}\nopagebreak

  \item grounding substitution~~\dotfill 5

  \indexspace
{\bfseries\hfil\normalsize I\hfil}\nopagebreak

  \item isolated form~~\dotfill 7

  \indexspace
{\bfseries\hfil\normalsize K\hfil}\nopagebreak

  \item Knuth-Bendix order~~\dotfill 3

  \indexspace
{\bfseries\hfil\normalsize N\hfil}\nopagebreak

  \item $\nat $ --- the set of natural numbers~~\dotfill 2

  \indexspace
{\bfseries\hfil\normalsize P\hfil}\nopagebreak

  \item precedence relation~~\dotfill 2

  \indexspace
{\bfseries\hfil\normalsize Q\hfil}\nopagebreak

  \item quasi-flat term~~\dotfill 12

  \indexspace
{\bfseries\hfil\normalsize R\hfil}\nopagebreak

  \item row~~\dotfill 8

  \indexspace
{\bfseries\hfil\normalsize S\hfil}\nopagebreak

  \item satisfiability~~\dotfill 4
  \item signature~~\dotfill 2
  \item simple~~\dotfill 7
  \item size~~\dotfill 8
  \item solution~~\dotfill 5
  \item substitution~~\dotfill 5
    \subitem grounding~~\dotfill 5

  \indexspace
{\bfseries\hfil\normalsize T\hfil}\nopagebreak

  \item $\TA (\Sigma )$~~\dotfill 2
  \item $\TA ^+(\Sigma )$~~\dotfill 4
  \item term algebra~~\dotfill 2
  \item term algebra sort~~\dotfill 4
  \item $\TT $~~\dotfill 18
  \item triangle form~~\dotfill 7

  \indexspace
{\bfseries\hfil\normalsize V\hfil}\nopagebreak

  \item validity~~\dotfill 4

  \indexspace
{\bfseries\hfil\normalsize W\hfil}\nopagebreak

  \item $w$ --- weight function~~\dotfill 2
  \item weight~~\dotfill 2
    \subitem of function symbol~~\dotfill 2
  \item weight function~~\dotfill 2
  \item working constraint~~\dotfill 8

\end{theindex}

\end{small}


\end{document}